\documentclass[12pt]{article}

\usepackage{graphicx}
\usepackage{ifthen}

\usepackage{amssymb}
\usepackage{amsmath}

\usepackage{amsthm}
\theoremstyle{plain}

\usepackage{bm}

\usepackage{chngcntr}
\usepackage{enumitem}

\usepackage{bigints}

\usepackage{color}
\usepackage[usenames,dvipsnames,svgnames,table]{xcolor}

\usepackage{bbm}

\usepackage{hyperref}
\usepackage{ulem}

\usepackage[ruled]{algorithm2e} 

\newcommand{\mket}[1]{| #1 \rangle}
\newcommand{\mbra}[1]{\langle #1 |}
\newcommand{\mbraket}[2]{\langle #1 | #2 \rangle}

\newcommand{\bI}{\mathbb{I}}
\newcommand{\bC}{\mathbb{C}}
\newcommand{\bH}{\mathbb{H}}
\newcommand{\bR}{\mathbb{R}}
\newcommand{\bU}{\mathbb{U}}
\newcommand{\bN}{\mathbb{N}}

\newtheorem{remark}{Remark}

\usepackage{color}
\usepackage{ulem}
\title{Variational Quantum Eigensolver for Classification in Credit Sales Risk}

\author{
	Joanna Wi\'sniewska$^{1, 0000-0002-2119-3329}$, \\ Marek Sawerwain$^{2, 0000-0001-8468-2456}$ \\ ~ \\
	{\small ${}^{1}$Institute of Information Systems, Faculty of Cybernetics,} \\
	{\small Military University of Technology,  Gen.~S.~Kaliskiego 2,} \\ {\small Warszawa 00-908, Poland, email: JWisniewska@wat.edu.pl} \\
	{\small ${}^{2}$ Institute of Control \& Computation Engineering} \\
	{\small University of Zielona G\'ora, Licealna 9,} \\ {\small Zielona G\'ora 65-417, Poland, email: M.Sawerwain@issi.uz.zgora.pl}  
}

\begin{document}
	
\maketitle
	
\begin{abstract}
The data classification task is broadly utilized in numerous fields of science and it may be realized by different known approaches (e.g. neural networks). However, in this work, quantum computations were harnessed to solve the problem. We take into consideration a quantum circuit which is based on the Variational Quantum Eigensolver (VQE) and so-called SWAP-Test what allows us to solve a classification problem connected with credit sales. More specifically, we cope with a decision problem of determining customer's reliability based on values of selected decision variables (e.g. generated turnover, history of cooperation). 

The classical data samples are converted into normalized quantum states. After this operation, samples may be processed by a circuit of quantum gates. The VQE approach allows training the parameters of a quantum circuit (so-called ansatz) to output pattern-states for each class. In the utilized data set, two classes may be observed -- cases with low and high credit risk. However, the VQE circuit differentiates more classes than two (introduces more detailed cases) and the final results are obtained with the use of aforementioned SWAP-Test. The elaborated solution is compact and requires only logarithmically increasing number of qubits (due to the exponential capacity of quantum registers).

Because of the low complexity of the presented quantum circuit, it is possible to perform experiments on currently available quantum computers, including the Noisy Intermediate-Scale Quantum (NISQ) devices. This type of devices, despite the presence of noise, is capable of solving the task analyzed in this work. All calculations, simulations, plots, and comparisons were implemented and conduced in the Python language environment. Source codes for each example of quantum classification can be found in the source code repository.  

	{\bf Keywords:} variational quantum algorithms, classification, credit risk analysis, quantum circuits
\end{abstract}

\section{Introduction} \label{lbl:sec:introduction:ec:2024:jw:ms}

Nowadays, the Noisy Intermediate-Scale Quantum (NISQ) devices \cite{Preskill2018} are regarded as a very promising direction of quantum computing development. The NISQ technology, due to limited number of qubits and known error correction schemes, may be successfully harnessed as a hardware for the Variational Quantum Algorithms (VQA) approach \cite{Wecker2015}, \cite{Cerezo2021VQA}, \cite{Zoratti2023}. This group of solutions assumes that there are issues which may be easily solved by classical computers and other problems which may be tackled with quantum computing. The VQA is a hybrid method which utilizes structurally uncomplicated quantum circuits for solving different problems and classical computers for optimization of mentioned circuits' parameters. The VQA method became so popular that we can observe various subtypes of it, e.g. the Quantum Approximate Optimization Algorithm (QAOA) \cite{Farhi2014} and the Variational Quantum Eigensolver (VQE) \cite{Peruzzo2014}, \cite{McClean2016}, 
\cite{Rattew2019}, \cite{Chivilikhin2020}, \cite{Fernandez2023},  \cite{Cerezo2022}, \cite{Fedorov2022}, \cite{Tilly2022} which is also utilized in this work. A quantum circuit in the VQE calculates the eigenvector correlated with the lowest eigenvalue of a ground state Hamiltonian. This method have found its application in many problems, e.g. drug discovery \cite{Blunt2022}, material science \cite{Lordi2021}, chemical engineering \cite{Cao2019}. Also the SWAP-Test \cite{Barenco1997}, i.e. a method which we utilize to compare two qubit states, is an another example of a low complexity circuit, what makes it easy to realize with the NISQ devices.

The VQE approach may be applied to point out some particular cases in a class to which we want to classify a data sample. In our data set, we can extract two main classes, but there are additional subgroups of samples. The VQE circuit realizes the learning process and indicates individual cases within a given class. Then, the SWAP-Test compares a data sample with other observations from the same class.

Therefore, our main aim is to use a combined solution -- VQE and SWAP-Test -- to the classification problem. Training data utilized in this work was gained from one of Polish pharmaceutical wholesalers and concerns some period of time in 2003.     

The paper is organized as follows: in Sec.~\ref{lbl:sec:preliminaries:EC:2024:jw:ms} preliminary information about notation is given. We also give briefly sketch and outline information about quantum computations, especially we give definition about a quantum register notion and a elementary information about unitary and measurement operation. In Sec.~\ref{lbl:sec:desision:model:ec:2024:jw:ms} a decision problem is presented. It is shown that there are different types of variables with various ranges in the analyzed data set, so the normalization process has to be performed. There are also presented the results of classification with the use of classical models which are later compared with the output of the {VQE+SWAP-Test method}. We also analyze an entanglement presence in the used quantum register, because it influences a form of the ansatz circuit constructed in the VQE approach. The proper form of the ansatz circuit, i.e. a shorter form, naturally enables faster simulations and shortens the process of learning. The limited depth of quantum circuits is very important to obtain a better accuracy for NISQ devices performing such circuits. Sec.~\ref{lbl:sec:class:with:vqe:technique:iccs:2023:jw:ms} is dedicated to the classification method based on the VQE approach and the SWAP-Test. There is also presented utilized quantum circuit at Sec.~\ref{lbl:vqe:for:class:and:analysis:iccs:2023:jw:ms} and its effectiveness calculated during a numerical simulation. The conclusions are situated in Sec.~\ref{lbl:sec:conclusions:iccs:2023:jw:ms}. The acknowledgments and bibliography are the last elements of this article.

\section{Preliminaries} \label{lbl:sec:preliminaries:EC:2024:jw:ms}

In this section, we would like to define some terms from the field of quantum computing, which we use in our work. Further information on this topic can be found in \cite{Nielsen2010}. First of all, the definition of quantum bit, so-called qubit, is needed. This basic unit quantum information may be presented as a two-element vector:

\begin{equation}
\mket{\psi}=\left[
\begin{array}{c}
\alpha _1 \\
\alpha _2
\end{array}
\right],
\end{equation}
under the normalization condition:
\begin{equation}
\left|\alpha _1\right|^2+\left|\alpha _2\right|^2=1, \; \; \alpha_1,\alpha_2 \in \mathbb{C}.
\label{iccs2023:norm}
\end{equation}
The marking $\mket{\psi}$ means the column vector, named $\psi$, in the Dirac notation. Another way to express the state of a qubit is the base-dependent superposition equation:
\begin{equation}
\mket{\psi}=\alpha _1 \mket{0} + \alpha _2 \mket{1},
\end{equation} 
where also the normalization condition (\ref{iccs2023:norm}) must be fulfilled. The vectors $\mket{0}$ and $\mket{1}$ span the computational basis which is called the standard basis:
\begin{equation}
\mket{0}=\left[
\begin{array}{c}
1 \\
0
\end{array}
\right], \mket{1}=\left[
\begin{array}{c}
0 \\
1
\end{array}
\right].
\end{equation}
The number of, possible to designate, bases is infinite -- the basis for one-qubit has to contain two orthonormal (orthogonal and normalized) vectors. Two vectors, e.g. $\mket{0}$ and $\mket{1}$, are orthogonal if their inner product $\mbraket{0}{1}=\mbraket{1}{0}=0$, and the vector $\mbra{0}$ is the complex conjugate of $\mket{0}$.

Furthermore, one qubit is not very useful if we want to perform any computation. The system of many qubits is called a quantum register. If we have $n$ qubits and want to join them into the register, we have to perform the operation of tensor product on the consecutive units, e.g. for two normalized vectors $\mket{\psi}$ and $\mket{\phi}$:
\begin{equation}
\mket{\psi}=\left[
\begin{array}{c}
\alpha _1 \\
\alpha _2
\end{array}
\right],
\mket{\phi}=\left[
\begin{array}{c}
\beta _1 \\
\beta _2
\end{array}
\right],
\mket{\psi} \otimes \mket{\phi}=\left[
\begin{array}{c}
\alpha _1 \mket{\phi} \\
\alpha _2 \mket{\phi}
\end{array}
\right]=\left[
\begin{array}{c}
\alpha _1 \beta _1 \\
\alpha _1 \beta _2 \\
\alpha _2 \beta _1 \\
\alpha _2 \beta _2
\end{array}
\right] = \mket{\psi \phi}.
\end{equation}  
We can also express this operation by the superposition equation:
\begin{multline}
\mket{\psi} \otimes \mket{\phi}=(\alpha _1 \mket{0} + \alpha _2 \mket{1}) \otimes (\beta _1 \mket{0} + \beta _2 \mket{1}) = \\
\alpha _1 \beta _1 \mket{00} + \alpha _1 \beta _2 \mket{01} + \alpha _2 \beta _1 \mket{10} + \alpha _2 \beta _2 \mket{11} = \mket{\psi \phi}.
\end{multline}
As we can see, the number of elements vital to describe $n$-qubit state is $2^n$, and we call them amplitudes. It is the number of vector elements or/and the number of components in the superposition equation. Of course, many-qubit states have to be normalized, so their amplitudes $\alpha_i$, which are complex numbers, must follow:  
\begin{equation}
\sum_{i=1}^{2^n} \left|\alpha_{i}\right|^2 = 1.
\label{ec2024:norm:n}
\end{equation}

There are two main types of actions which may be carried out on quantum states: computational operation and measurement. 

The computational operations are performed by so-called quantum gates which may be represented by unitary matrices. This unitarity feature allows keeping the normalization of quantum states after the operation. The mathematical operation realized here is multiplying a unitary matrix $U \in \bU(\bC^d)$ (where $\bU$ represents a whole set of unitary operators acting in the selected space $\bC^d$) by a vector state $\mket{\psi}$:
\begin{equation}
U \mket{\psi} = \mket{\psi'}, 
\end{equation}
where $U$ is sized $2^n \times 2^n$ if it transforms $n$-qubit state. It should be also noted that in this paper $U$ operators act on the pure states $\partial E(\bC^d)$. The output state $\mket{\psi'}$ is of the same size as $\mket{\psi}$.

Of course, after a series of computations on a quantum state, one would like to know its result, so the measurement operation have to be performed. In this case, the von Neumann approach may be utilized:
\begin{equation}
\mket{\psi'} = \frac{P_i \mket{\psi}}{\sqrt{\mbra{\psi} P_i \mket{\psi}}}, 
\end{equation}
where the measured state $\mket{\psi'}$ is calculated with a use of the initial state $\mket{\psi}$ and the projection operator $P_i$. The von Neumann's quantum measurement is performed in a chosen computational basis. It is about projecting the whole state $\mket{\psi}$, or its particular qubits, to the basis states. Naturally, it is only a mathematical model of physical experiments, where $P_i$ describes one of all possible projections, e.g. in the SWAP-Test, there is a three-qubit circuit and the probability of measuring $\mket{0}$ on the first qubit is crucial, so in this case $P_i=P_0 \otimes \bI \otimes \bI$, where $P_0=\mket{0}\mbra{0}$ projects the first qubit into $\mket{0}$ from the standard basis and other two qubits are not measured (what is marked as the identity operation $\bI$). 

To summarize, utilized in this work, notations and symbols, we gathered all the most important denotations at the Table~\ref{lbl:tbl:notations:table:ec:2024:jw:ms}.
	
\begin{table}
	\caption{Symbols and notations used in the paper}
	\label{lbl:tbl:table:of:symbols}
	\begin{center}
		\begin{tabular}{c|l}
			\hline\hline
			Notation & Description \\ \hline\hline
			$\bN$, $\bR$, $\bC$ & set of integer, real, complex numbers \\
			$\bH$ & Hilbert space \\
			$\alpha$, $\beta$, $\gamma$, $\lambda$, $\ldots$ & real or complex numbers \\
			$i,j,k,l$ & integer numbers used for indexing \\
			$\otimes$, $\times$ &  \parbox{9.5cm}{tensor product and multiplication or product of two scalars and/or matrices} \\
			\hline
			$\bI$, $\bI_N$, $\bI_{d^N}$ & \parbox{9.5cm}{identity matrix/operator, identity operator defined in a system with $N$ qubits, identity operator with declared dimensionality} \\ 
			$\bU(\bC^d)$  &  group of unitary matrices acting in $\bC^d$ \\ 
			$U$, $U_{(i)}$ & \parbox{9.5cm}{general unitary operator $U$, and operator $U_{(i)}$ applied to the i-th qubit/qudit in a quantum register} \\ \hline
			$A^{\star}$, $A^{\dagger}$ & conjugation of matrix, Hermitian adjoint of matrix \\ 
			$\mbraket{ - }{ - }$ & inner (scalar) product \\
			$\partial E(\bC^d)$ &  set of pure states on $\bC^d$ \\ \hline
			$1\! \ldots \!n$ & means the sequence of $1, 2, 3,\ldots, n$ \\
			\hline
			$X$, $Y$ & $X$ input vectors data set and $Y$ binary output (0/good, 1/bad)\\
			$(x_1,x_2,x_3,x_4,x_5,x_6,x_7)_{i}$ & i-th input data vector \\
			$D$ &  dataset for classification\\
			$c$, $g^c$ & class number, group number in class c\\
			$c_n$, $c^{g}_{n}$ & \parbox{9.5cm}{the number of main classes, and the number of groups in class $c$} \\
			$\theta_{T^g_c}$ & parameters of the ansatz circuit for class $c$ and group $g$ \\
			$(\Lambda^g_c, \Omega^g_c)$ & the acceptance range for given group $g$ in class $c$ \\
			\hline 
			$T_{ini}$, $T_{VQE}$, $T_{SWAP}$, $T_{qtot}$ & \parbox{9.5cm}{the computational complexity of, respectively, initial, VQE, SWAP, and the whole quantum circuit} \\
			\hline
			$\mket{\Psi}$, $\mket{\psi_{ans}}$, $\mket{\psi_{VQE}}$,  $\mket{\psi_i}$ & the whole quantum register and subregisters \\
			$H$ & Hamiltonian describing system's energy\\
			$E_0$ & ground state energy of $H$\\ 
			$E_{VQE}$ & energy of VQE system\\ 
			\hline\hline
		\end{tabular}
	\end{center}
	\label{lbl:tbl:notations:table:ec:2024:jw:ms}
\end{table}

\section{Decision Situation Model} \label{lbl:sec:desision:model:ec:2024:jw:ms}

In this paper, we deal with a decision task focusing on a risk assessment of some credit sales -- this example was already considered in works \cite{Wisniewska2009, WisniewskaSawerwain2020}. The adopted business model assumes that wholesalers cooperate with some companies which order offered products. The wholesaler delivers goods, issues an invoice, and the customer is obliged to pay within the period specified on the invoice. Sometimes, of course, customers place new orders even if they have overdue invoices. At this moment, employees of the wholesaler's debt collection department should react -- contact with the client and make a decision if the orders may be realized or should be suspended. We assume that the factors which should be taken into consideration in this decision situation are:
\begin{enumerate}
	\item[(a)] the amount of the customer's arrears (real number),
	\item[(b)] the amount of all current dues (real number),
	\item[(c)] the number of days relating to the payment delay of the oldest overdue invoice (integer),
	\item[(d)] the payment amount declared as already issued, but still not posted on the account (real number),
	\item[(e)] customer's reliability in the past cooperation (Boolean value),
	\item[(f)] evaluation of customer's cooperating entities (Boolean value),
	\item[(g)] a promissory note from the customer (Boolean value).
\end{enumerate}	

If a customer has overdue invoices then the case is forwarded to the debt collection department where an employee has to make a decision if the orders' realization should be stopped. A mathematical model may be presented as two vectors $X$ and $Y$ -- the first containing the input data according to seven factors mentioned above (the i-th variable $x_i$ corresponds to the i-th factor) and the second vector with binary output (0 -- the goods may be issued, 1 -- otherwise (sale blocking)):

\begin{equation}
X=(x_1,x_2,x_3,x_4,x_5,x_6,x_7), \; Y=(y_1), \; y_1\in\{0,1\}.
\end{equation}  

The data set, used in this work, was constructed with the use of some practical directions, e.g. if the amount of the customer's arrears exceeds 20\% of the whole amount of all current dues, the case is threatened with a sales ban. Another alarming factor is more than 6 days of delay in paying of the oldest due. Mitigating circumstances are: declared payment, positive past experience in the cooperation with the client, good opinion about client's recipients, promissory notes which could make easier potential debt recovery process.

\subsection{Data Normalization Processes} \label{lbl:sec:data:preparation:iccs:2023:jw:ms}

Before we will be able to start a learning process, which is the process of fitting the quantum circuit's parameters, it is important to prepare data. The training set contains variables of different types and ranges, so avoiding the data normalization process could let to incorrect influence of some variables on obtained results, e.g. one variable accepts values in the range [0; 10000] and another obtains only Boolean values.

We utilize two steps in data normalization: classical and quantum. Firstly, we follow the widely known normalization procedure to convert all values to the range [0; 1] (of course, if we deal with Boolean variables, this is redundant). Let us assume that $i$ stands for the index of a variable and the number of variables is $a$, so $i=1, 2, ..., a$. Consequently, $j$ is the index of an observation from the data set and $b$ is the number of observations ($j=1, 2, ..., b$). Now, for each i-th variable, the normalization process have to be done over all observations. The normalized values $\tilde{x}_{i,j}$ are calculated according:
\begin{equation}
	\tilde{x}_{i,j}=\frac{x_{i,j}-\min\limits_{1 \leq j \leq b}\{x_{i,j}\}}{\max\limits_{1 \leq j \leq b}\{x_{i,j}\}-\min\limits_{1 \leq j \leq b}\{x_{i,j}\}}.
\end{equation} 
At this moment, the variables are normalized but the quantum circuit, which solves the problem, needs an input in a form of correct quantum states. So the second, quantum, step of normalization has to be realized.

Quantum states of $n$ qubits have $2^n$ amplitudes $\alpha_i$ because each pure quantum state may be described as a superposition of the standard basis states $\mket{i}$: 
\begin{equation}
\mket{\psi} = \sum_{i=0}^{2^n-1} \alpha_i \mket{i},
\end{equation}
where $i$ are encoded as decimal numbers but some times, e.g. to describe standard basis vectors, binary numbers are used. Amplitudes $\alpha_i$ are complex numbers and the normalization condition holds:
\begin{equation}
\sum_{i=0}^{2^n-1} |\alpha_i|^2 = 1 .
\end{equation}

If we have $a$ variables, the size $n$ of a utilized quantum register has to follow the rule:
\begin{equation}
	a \leq 2^n.
\end{equation}   
If $a = 2^n$ then all amplitudes have ensured values. If $a < 2^n$, what is more probable, missing $c$ variables ($a+c = 2^n$) have to be added to the data set with zero values, e.g. in our problem with credit sales, we have seven variables, so the eight variable $x_8$ has to be introduced and in each observation the value of this variable is equal zero.

Finally, data has to be converted to quantum states, what may be realized by changing values of variables within every j-th observation: 
\begin{equation}
\hat{x}_{i,j}=\sqrt{\frac{\tilde{x}_{i,j}}{\qquad \sum_{i=1}^{2^n} \tilde{x}_{i,j}}}.
\label{jw:ms:iccs2023:norm}
\end{equation} 
Now, observations fulfill the normalization condition for quantum states -- the variables' values $\hat{x}_{i,j}$ of a $j$-th observation became amplitudes' values $\alpha_i$ of a $j$-th quantum state.

\begin{remark}
The normalization process in the classical (not quantum) machine learning helps to limit a too strong influence of some variables on the model's parameters. The same effect we would like to obtain in our calculations, so the two steps of normalization are performed -- classical and quantum (to provide correct quantum states). 
\end{remark}
	
In the following sections of this work, we utilize the data set which is normalized to the form of quantum states. Firstly, we performed classical normalization on variables $x_1$ to $x_4$ because they were not Boolean values. Then, we added variable $x_8$ with zero values to make sure that we supply values for all amplitudes in a three-qubit state ($n=3$). In the last step, we complete the normalization process in the meaning of (\ref{jw:ms:iccs2023:norm}).

\subsection{Entanglement in Samples} \label{lbl:ssec:entanglement:in:probes:iccs:2023:jw:ms}

After the double normalization process, we obtain samples in the form of quantum states, which are build upon the observations from the data set. A quantum state $D_j$ corresponds to the $j$-th observation. The analyzed example needs three qubits, so:
\begin{equation}
	\mket{D_j} = \mket{q_0 q_1 q_2} = \sum_{k=0}^{7} \alpha_k \mket{k}.
\end{equation}
The number of variables in our example is seven, so the last amplitude is always zero. The analysis of the entanglement's presence was carried out for both classes (marked as zero when goods may be issued and one when sales should be blocked) with the use of the EntDetector package \cite{EntDetector2021}. We can observe the entanglement between qubits specified in round brackets:
\begin{equation}
	\mathrm{class} \; 0\!: \; (0,1,2), \;\;\; (1, 2) \;\;\; \mathrm{class} \; 1\!: \; (0,1,2), \;\;\; (0, 2), \;\;\; (1,2).
\end{equation}
In the most cases all three qubits are entangled. There are only four samples in class 0 where only qubits (1, 2) are entangled. In the class 1, there are four cases of entangled (0, 2) and eight incidents for (1, 2). The entanglement was diagnosed by the method detection\_entanglement\_by\_paritition\_division, from the EntDetector package, which is based on the partial trace calculation. 

\begin{remark}
The presence of entanglement in the data means that in the further part of the classification process, using the VQE method, the ansatz circuit must correctly recreate entanglement in accordance with the data samples.	  
\end{remark}

\subsection{Classical Methods in the Analysis of Training Data Set}
	
Before the results of quantum approach will be presented, we would like to pay some attention to classical methods of machine learning which may be applied in the case of analyzed problem. 

\begin{figure}
	\includegraphics[width=\textwidth]{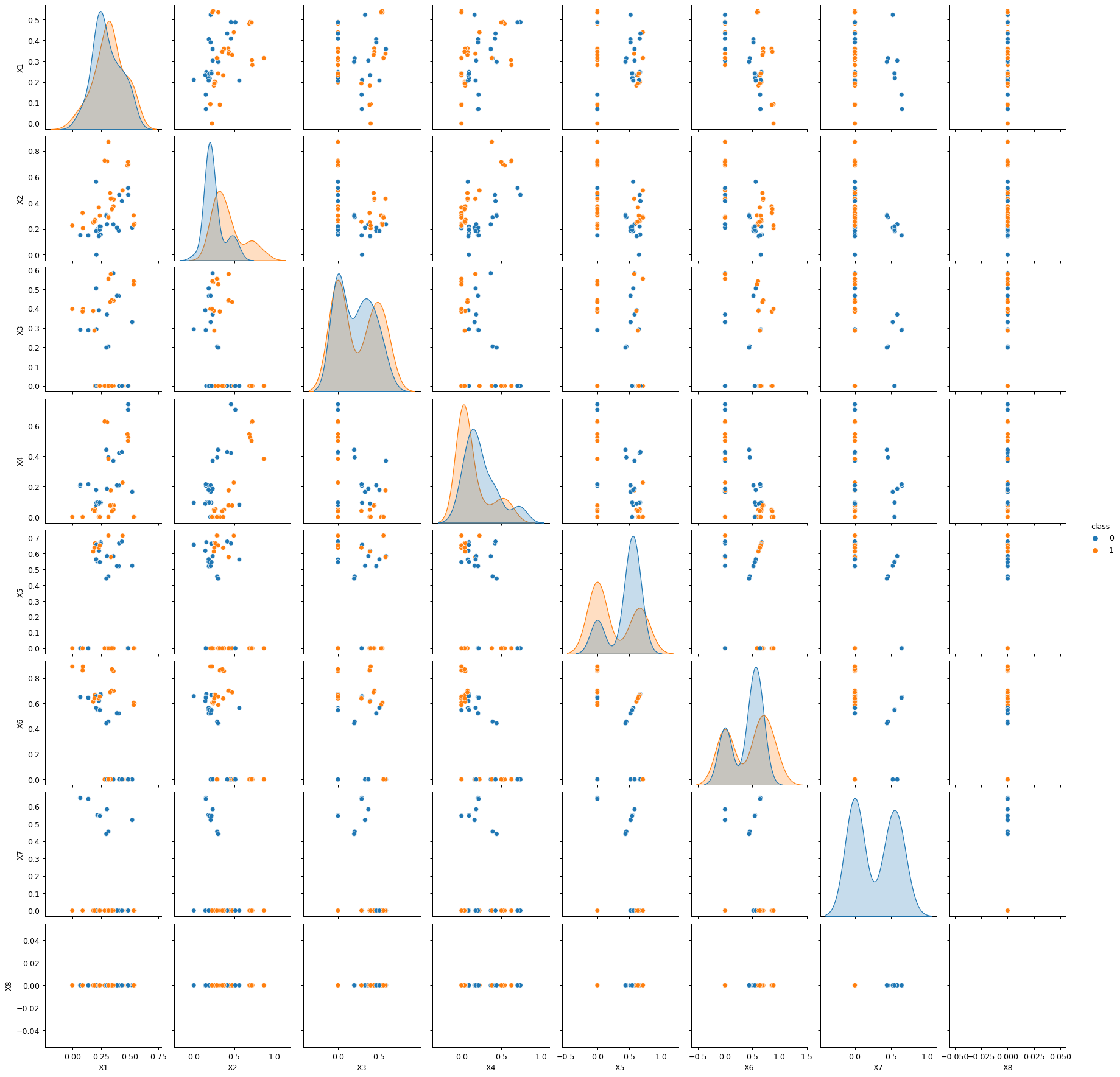}
	\caption{Plot generated by pair-plot function from the Seaborn library for Python which depicts class separation for each pair of variables. Results for the variable $x_8$ should be omitted because values of this variable are always zero (auxiliary variable added during the normalization process)}
	\label{lbl:fig:pair:plot:training:data}
\end{figure}

It is easy to show that the pair plot method from the Seaborn library \cite{Waskom2021} for Python indicates that the classes in our problem are not well separated (there are no consistent regions formed by one color). As it is shown in Fig.~\ref{lbl:fig:pair:plot:training:data}, the cases when the cooperation with the client was continued are marked with the blue color and an orange color symbolizes the sale blockade.  The variable $x_7$ is a small exception because there are some regions of blue points, but it does not change the fact that it is hard to state the visible partition to classes in the two-dimensional plot.
Correlations between variables are weak, what was tested with the use of the correlation matrix containing Pearson coefficients. Table~\ref{lbl:tbl:pearson:coefficients}, presenting obtained results (for the variable $x_8$ it is impossible to calculate values of coefficients because its constant character), is available at our source code repository \cite{VQEClassRepo}.

\begin{table}[h!]
\caption{Pearson coefficients for normalized variables in the analyzed problem}
\begin{center}
\begin{tabular}{r|c|c|c|c|c|c|c} 
  & $x_1$ & $x_2$ & $x_3$ & $x_4$ & $x_5$ & $x_6$ & $x_7$ \\
\hline
$x_1$ &  1.000000 &  0.395159 &  0.002715 &  0.419183 & -0.157285 & -0.543081 & -0.281667 \\
$x_2$ &  0.395159 &  1.000000 & -0.392805 &  0.560917 & -0.437261 & -0.504038 & -0.403142 \\
$x_3$ &  0.002715 & -0.392805 &  1.000000 & -0.310900 &  0.056815 &  0.031139 & -0.134183 \\
$x_4$ &  0.419183 &  0.560917 & -0.310900 &  1.000000 & -0.305393 & -0.640078 & -0.053246 \\
$x_5$ & -0.157285 & -0.437261 &  0.056815 & -0.305393 &  1.000000 & -0.146336 &  0.091410 \\
$x_6$ & -0.543081 & -0.504038 &  0.031139 & -0.640078 & -0.146336 &  1.000000 &  0.056590 \\
$x_7$ & -0.281667 & -0.403142 & -0.134183 & -0.053246 &  0.091410 &  0.056590 &  1.000000
\end{tabular}
\end{center}
\label{lbl:tbl:pearson:coefficients}
\end{table}

\begin{figure}[ht]
\begin{center}
\includegraphics[width=0.8\textwidth]{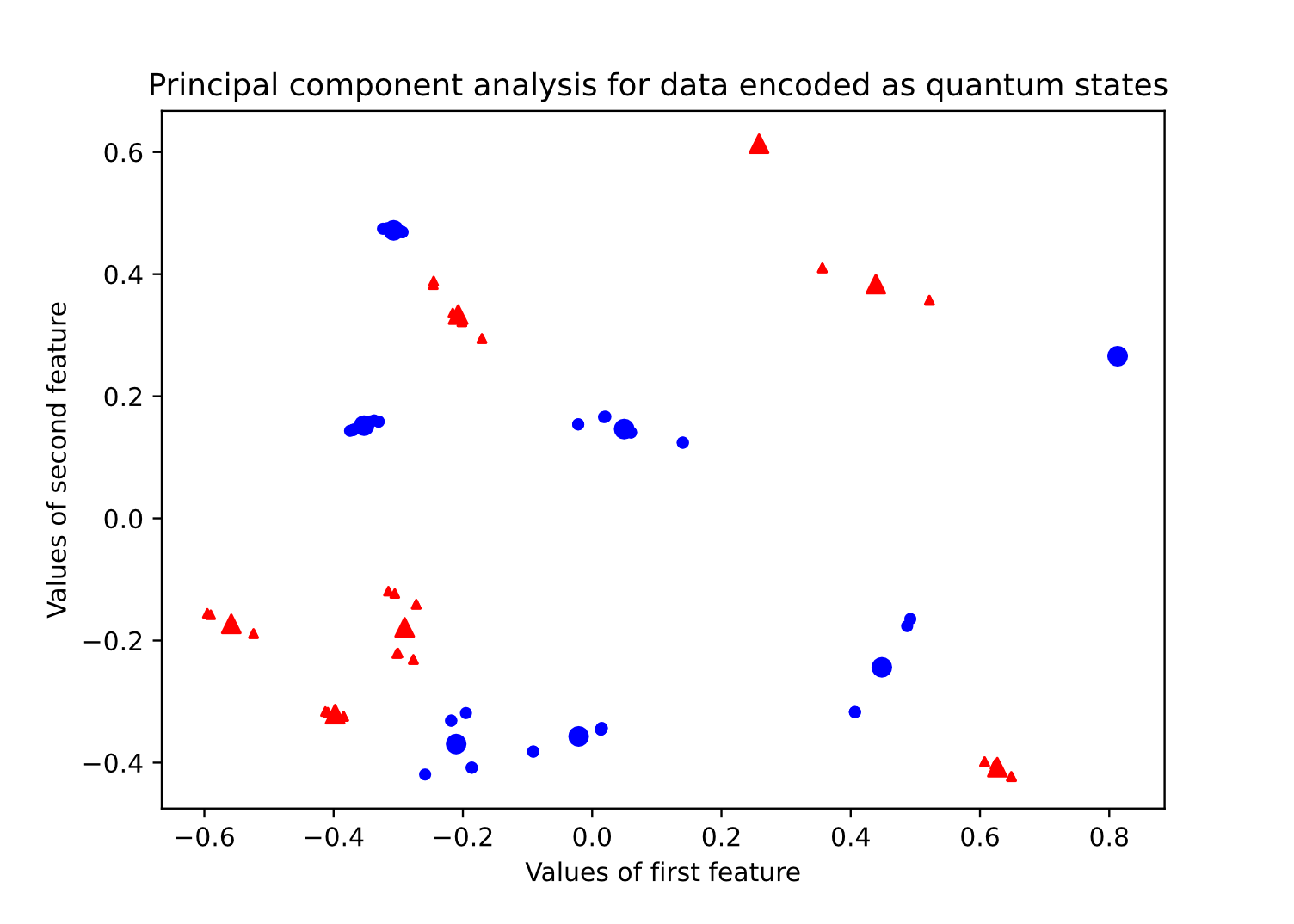}
\end{center}
\caption{Decomposition of the eight dimensional data (train data with 40 probes) to two dimensions with the use of PCA method from the Scikit-Learn \cite{sklearn_url} library for Python. The blue color (filled circle) is related to class 0 and the red one (triangles) to class 1. Each class has been divided into seven clusters. It is necessary to underline that some samples after the PCA are very close to one another and the figure does not illustrate them}
\label{lbl:fig:pca:decomposition:two:comp:training:data}
\end{figure}

At this point, we can claim that there is no simple linear model which fits to our data. Anyway, other models may be used. Finishing the phase of the input data exploration, we find the Principal Component Analysis (PCA) as interesting. It allows to lower the number of dimensions in our problem of eight variables with the Singular Value Decomposition (SVD). Fig.~\ref{lbl:fig:pca:decomposition:two:comp:training:data} depicts obtained results. It may be observed that there are clusters formed by observations from the same classes. 

\begin{remark}
All scripts generating Seaborn plots, Fig.~\ref{lbl:fig:pca:decomposition:two:comp:training:data}, and data for Pearson coefficients shown in Table~\ref{lbl:tbl:pearson:coefficients}  are available in the source code repositories \cite{VQEClassRepo}. In addition, repositories contain all source codes for numerical experiments conducted and discussed in this paper.
\end{remark}

\subsection{Accuracy of classical classifiers}
\label{lbl:sec:accuracy:of:classical:techniques:iccs:2023:jw:ms}

To evaluate the accuracy of quantum approach, it would be reasonable to compare its results with outcomes which may be received by non-quantum classifiers. Table~\ref{lbl:tbl:cc:accuracy} shows the obtained accuracy for different models. All calculations were realized with the use of procedures from the Scikit-Learn library \cite{sklearn_url}. The classifiers marked as SVM (Support Vector Machine) were simulated by the method SVC (Support Vector Classification) which is suitable for tasks with two and more classes. Tests were performed for different types of a kernel function (parameter k). Next used method is the k-nearest neighbours approach with the parameter n as the number of neighbours. In case of the rest of applied classifiers, their names are the same as names of the functions in the Scikit-Learn library and the parameters are shown in Table~\ref{lbl:tbl:cc:accuracy}.

It may be observed that some classifiers reached 100.0 \% fit to our training data. However, these cases present complex models with the polynomial kernel functions. We expect that the simulation of the hybrid model VQE is able to solve the task with the use of less complicated model.  


\begin{table}
\caption{Efficiency of classical classifiers for the problem of credit sales. The points from each class are spread along the space and do not form uniform clusters (Fig.~\ref{lbl:fig:pca:decomposition:two:comp:training:data}), what requires applying methods with polynomial kernel function to obtain high fit of the model. The word "default" in the column "Basic parameters" means that none additional parameters during the call of the given method have not been used.}
\begin{center}
\begin{tabular}{|l|l|c|}
\hline
Classifier short name			& Basic parameters 			& \parbox{2cm}{Accuracy for training set} \\ \hline \hline
			SVM 							& k=sigmoid, C=10			& 	62.5\%  				\\	\hline
			SVM 							& k=linear 					&   98.8\%  				\\	\hline
			SVM 							& k=rbf 					&   98.8\%  				\\	\hline
			SVM 							& k=poly 					&  100.0\%  				\\	\hline
			KNClassifier			  		& n=3 						&   98.8\%					\\	\hline
			GaussianProcessClassifier		& k = 1.0 * RBF(1.0)		&  100.0\% 				\\	\hline
			DecisionTreeClassifier			& max\_depth=5				&  100.0\% 				\\	\hline
			RandomForestClassifier			& \parbox{5cm}{max\_depth=5, n\_estimators=10, max\_features=8}	&  100.0\% 				\\	\hline
			MLPClassifier					& alpha=1, max\_iter=1000	&   98.8\% 				\\	\hline
			AdaBoostClassifier				& default					&  100.0\%  				\\	\hline
			GaussianNB						& default					&   75.0\% 				\\	\hline
			QuadraticDiscriminantAnalysis	& default					&   50.0\% 				\\	\hline
\end{tabular}
\end{center}	
\label{lbl:tbl:cc:accuracy}
\end{table}

\section{Classification based on the VQE and SWAP-Test circuits}  \label{lbl:sec:class:with:vqe:technique:iccs:2023:jw:ms}

\begin{figure}[!ht]
	\begin{center}
		\includegraphics[width=0.85\textwidth]{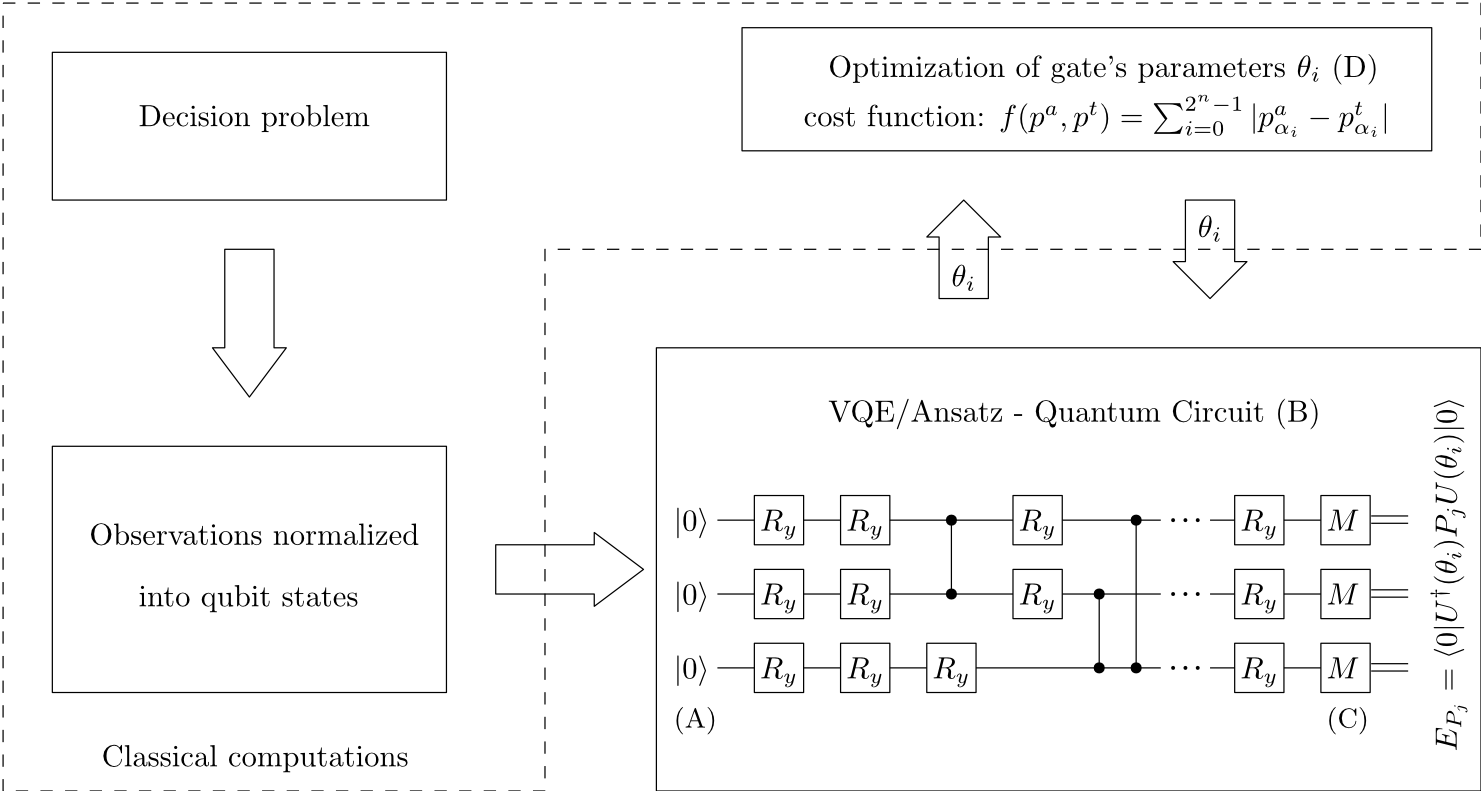}
	\end{center}
	\caption{A general diagram of a simplified form of the VQE method. The first stage (A) in the implementation of the VQE is determining the initial state $\mket{000}$. Then, the ansatz circuit (B) is implemented, which needs a series of parameters $\{ \theta_i \}$ which are determined by optimization using a classical machine. The circuit's task is to transform the input state into another form of the state that is subject to the measurement operation. After the measurement step (C), a value of the cost function (D) is calculated what is based on the obtained probability distribution. During the optimization process a value of the cost function and rotation parameters $\{ \theta_i \}$ are selected (e.g. using the gradient method) for gates. It is performed  so that the distribution obtained after the measurement (C) is as similar as possible in terms of the cost function to the desired probability distribution, representing, for example, a specific class of the sample under examination. The optimization process is carried out by a classical machine, but naturally the execution of the circuit is a task for a quantum machine}
	\label{lbl:fig:general:vqe:schema}
\end{figure}

The Variational Quantum Eigensolver (VQE) method was firstly proposed in \cite{Peruzzo2014} and further developed in \cite{McClean2016}. It aroused interest of many researchers and resulted in a lot of works, e.g. 
\cite{Cerezo2022}, \cite{Fedorov2022}, \cite{Tilly2022}. The VQE is included in the group of methods termed as Variational Quantum Algorithms (VQA) \cite{Wecker2015}, \cite{Cerezo2021VQA}, \cite{Zoratti2023} which are based on a hybrid solution combining quantum and classical computations.  The VQE method's scheme for the problem analyzed in this work is depict in Fig.~\ref{lbl:fig:general:vqe:schema} where we can observe two parts of the solution: data preparation and the optimization is realized on the classical computer, and so-called ansatz \cite{Wecker2015}, 
\cite{Nakaji2021}, 
\cite{Bilkis2023}, 
should be implemented as a quantum circuit. 

One of the VQE's advantages is that the quantum circuit may be realized on the NISQ devices \cite{Preskill2018}. This type of quantum machines is characterized by a fact that although the noise presence some computations still can be successfully performed.

The quantum circuit in the VQE produces an $n$-qubit state $\mket{\psi}$ which may be identified with the wave function of the same name. A ground state energy $E_0$ of the Hamiltonian $H$ is expressed as:  
\begin{equation}
E_0 \leq \frac{\mbra{\psi} H  \mket{\psi} }{ \mbraket{ \psi }{ \psi } }.	
\end{equation}
The main objective of the VQE is to determine the amplitudes of $\mket{\psi}$, such that an upper bound of $E_0$ will be found. It means that the expectation value of $H$ has to be minimized, because the lowest eigenvalue of $H$ is correlated to the sought eigenvector. This problem may be defined as determining a unitary operation $U$ with parameters $\{ \theta_i \}$ to fulfill the optimization problem:
\begin{equation}
E_{\mathrm{VQE}} =  \min_{\theta_i} \mbra{0} U^{\dagger} (\theta_i) H  U(\theta_i) \mket{0},
\label{lbl:eq:E:for:VQE}
\end{equation}
where $\mket{0}$ is the initial state such that $U(\theta_i) \mket{0}=\mket{\psi}$. In \cite{Tilly2022} the definition of the Hamiltonian $H$ is given as $H = \sum_j \omega_j P_j$, where $\omega_j$ are weights of $P_j$ which symbolize consequent Pauli operators: $X$, $Y$, $Z$, $I$ ($j=1, \ldots, 4$). It means that equation~(\ref{lbl:eq:E:for:VQE}) may be written as:
\begin{equation}
E_{\mathrm{VQE}} =  \min_{\theta_i} \sum_j \omega_j \mbra{0} U^{\dagger} (\theta_i) P_j U(\theta_i) \mket{0}.
\label{lbl:eq:E:for:VQE:with:P}
\end{equation}
Now, we can express an expectation value for each Pauli operator separately:
\begin{equation}
	E_{P_j} = \mbra{0} U^{\dagger} (\theta_i) P_j U(\theta_i) \mket{0} .
\end{equation}
The expectation values may be calculated with the use of a quantum device (e.g. NISQ). 

The optimization problem:
\begin{equation}
E_{\mathrm{VQE}} =  \min_{\theta_i} \sum_j \omega_j E_{P_j},
\label{lbl:eq:VQE:optimalization:process}
\end{equation} 
can be solved by a classical computer, using e.g. COBYLA or SPSA method from SciPy Python package \cite{Virtanen2020}. 

\begin{remark}
The VQE method is known for the presence of the Barren plateaus problem \cite{Kulshrestha2022} which appears during the optimization part. But, in examined data because of several smaller clusters, Barren plateaus problem has not been observed.	
\end{remark}

In this article, we utilize the VQE method to generate a state $\mket{\psi}$ which represent a class of analyzed observation. Moreover, the state obtained from the VQE is compared to the test state dedicated to a particular class. A distance between these vector states determines the class of an observation. Another important task is to implement the operation $U$ what is presented in Sec.~\ref{lbl:vqe:for:class:and:analysis:iccs:2023:jw:ms}.

Returning to the optimization problem, i.e. seeking of $E_{\mathrm{VQE}}$ value, it is naturally required to define the cost function which will be utilized during the whole optimization process. Let us mark it as $f(p^a, p^t)$ where $p^a$ stands for the description of a current quantum state, and $p^t$ is a final quantum state. The value of the cost function has to be minimized, as the difference between two mentioned states. A reasonable candidate for $f$ is the Fidelity measure \cite{Nielsen2010}. However, in this work, probability distributions of compared quantum states are used, and the sum of the absolute errors of particular probabilities is calculated as:  
\begin{equation}
	f(p^a, p^t) = \sum_{i=0}^{2^n-1} | p^a_{\alpha_i} - p^t_{\alpha_i}|,
	\label{lbl:eq:f:error:function:JW:MS:ICCS:2023}
\end{equation}
where $p^a_{\alpha_i}$ is the probability of obtaining state $\mket{i}$ after a measurement (what is given by the probability amplitude $\alpha_i$).

Such form of the cost function $f$ is determined by the way in which the information concerning probability distribution may be gained from the quantum device, e.g. IBM Q Experience, where repeating the experiment allows obtaining the probability distribution. This approach seems to be more realistic in a context of current capabilities of available quantum devices. Therefore, in the description of our experiment, we assume that the cost function $f$ depicts an error as the absolute difference between two probability distributions.    

\begin{figure}
	\includegraphics[width=\textwidth]{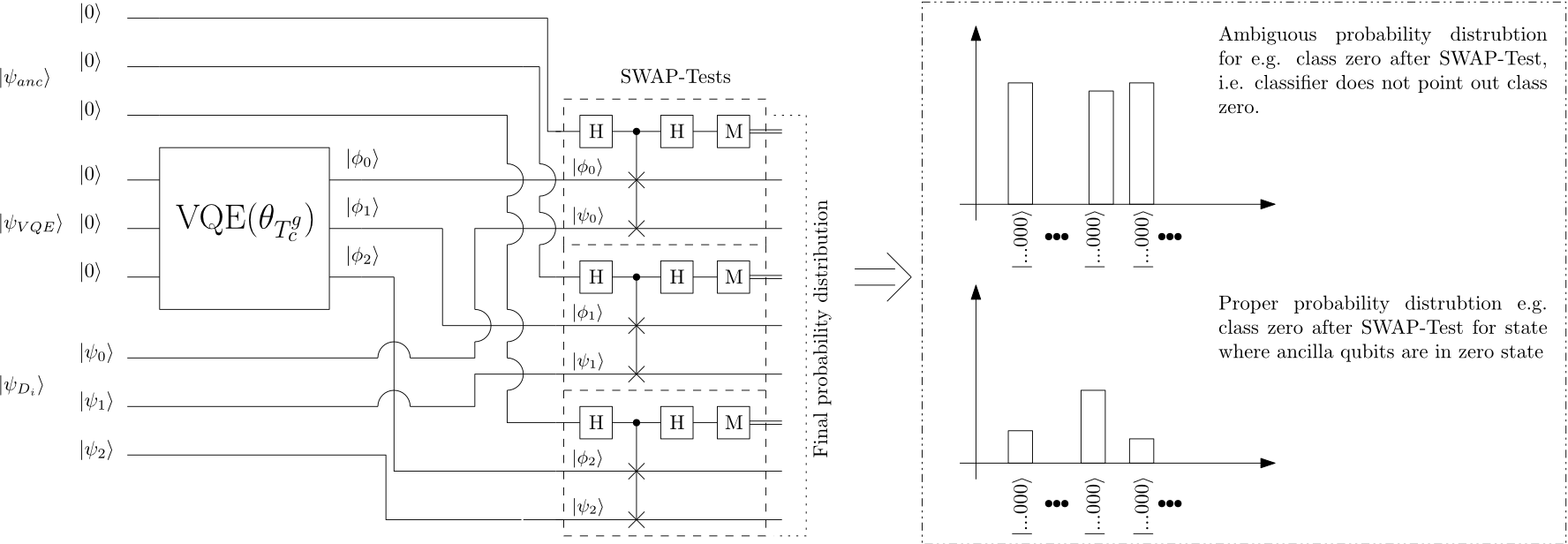}
	\caption{A generalized scheme of the classifying circuit for the VQE approach in case of a dataset discussed in this paper. The previously trained ansatz circuit generates pattern states according to its parameters $\theta_{T^{g}_c}$ what is unequivocal for a specific class. The SWAP-Test evaluates a distance between a sample and averaged pattern-state.}
	\label{lbl:fig:classification:with:VQE:JW:MW:ICCS:2023}
\end{figure}

It should be mentioned that quantum computing makes possible comparing two quantum states (e.g. $p^a$ and $p^t$) by so-called SWAP-Test \cite{Barenco1997}. It is possible because the SWAP-Test is based on the scalar product of examined states (in this aspect it is similar to the Fidelity measure). A general scheme of the SWAP-Test is also shown in Fig.~\ref{lbl:fig:classification:with:VQE:JW:MW:ICCS:2023}.   

\begin{figure}
	\begin{center}
		\includegraphics[height=2.5cm]{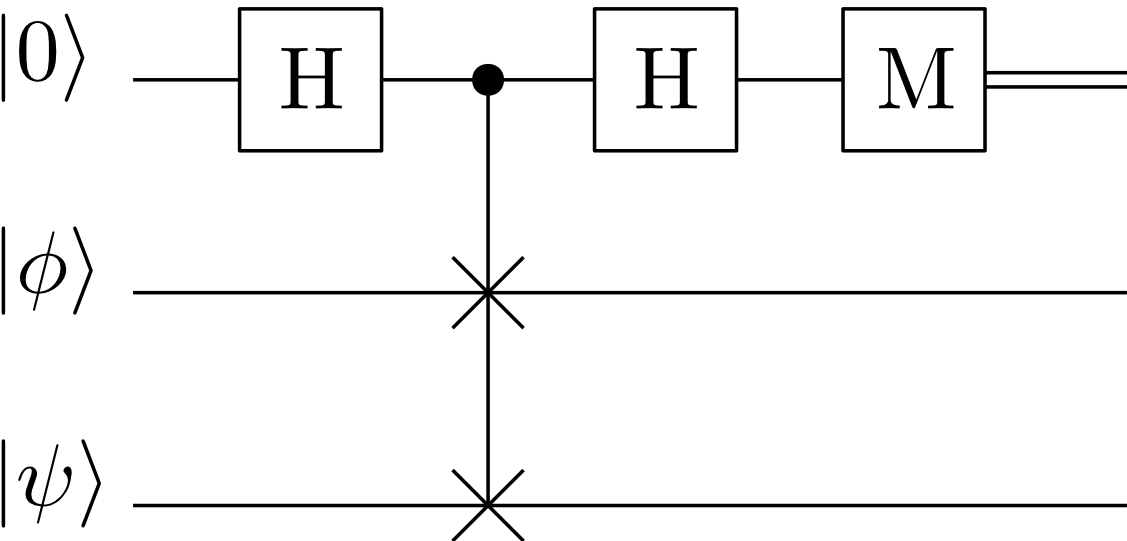}
	\end{center}	
	\caption{A scheme of the quantum circuit (two Hadamard gates represented by a square block with big letter H and Fredkin gate also called a controlled swap gate realizing the SWAP-Test ("overlap") between states $\mket{\phi}$ and $\mket{\psi}$. The input on the first qubit is always $\mket{0}$. The similarity of given states is contained in the first output qubit as the probability of measuring $\mket{0}$ (denoted as gate M) and it is equal to: $P_0 = \frac{1}{2} + \frac{1}{2}{|\mbraket{\psi}{\phi}|}^2$. If states are orthogonal (different) the probability is equal to 1/2 and is 1 if the states are equal}
	\label{lbl:fig:swap:circuit:JW:MS:ICCS:2023}
\end{figure}

This simple circuit allows measuring a state of the first output qubit. The resemblance of states $\mket{\psi}$ and $\mket{\phi}$ is reflected as the probability of obtaining state $\mket{0}$ on the first qubit is $P_0 = \frac{1}{2} + \frac{1}{2} {|\mbraket{\psi}{\phi}|}^2$.
If $P_0=1$ then analyzed states are the same, because it means that inner product $\mbraket{\psi}{\phi}=1$. The more states point out the opposite points on the Bloch sphere, the less similar they are. For the orthogonal states, the product $\mbraket{\psi}{\phi}=0$ and then $P_0=0.5$. The SWAP-Test is useful for checking the similarity of two quantum states. In addition, the SWAP-Test may be applied for each qubit utilized in the VQA method. An evaluation of the classification may be also realized as examining the probability distribution after the completed test. The observation is correctly classified if for each of three qubits the high probability of obtaining $\mket{0}$ is gained. The class is represented by the set of circuit's parameters $\theta_i$ in the VQE method.

\section{Variational Quantum Eigensolver for Credit Sales Risk Analysis} \label{lbl:vqe:for:class:and:analysis:iccs:2023:jw:ms}

The presented classification method with the use of the VQE requires preparation of the quantum circuit which in this approach is called an ansatz. This circuit is characterized in Sec.~\ref{lbl:ssec:ansatz:circuit:JW:MW:ICCS:2023}. The scheme of proceeding subsequent observations in the process of classification is shown in Sec.~\ref{lbl:ssec:alg:for:data:class:JW:MW:ICCS:2023}. The discussion about the results of the conducted experiments is included in Sec.~\ref{lbl:subsec:accuracy:JW:MS:ICCS:2023}.

\subsection{Creation of Ansatz Circuit} \label{lbl:ssec:ansatz:circuit:JW:MW:ICCS:2023}

\begin{figure}
\begin{center}	
	\begin{tabular}{c}
		(A) \\
		\includegraphics[width=0.80\textwidth]{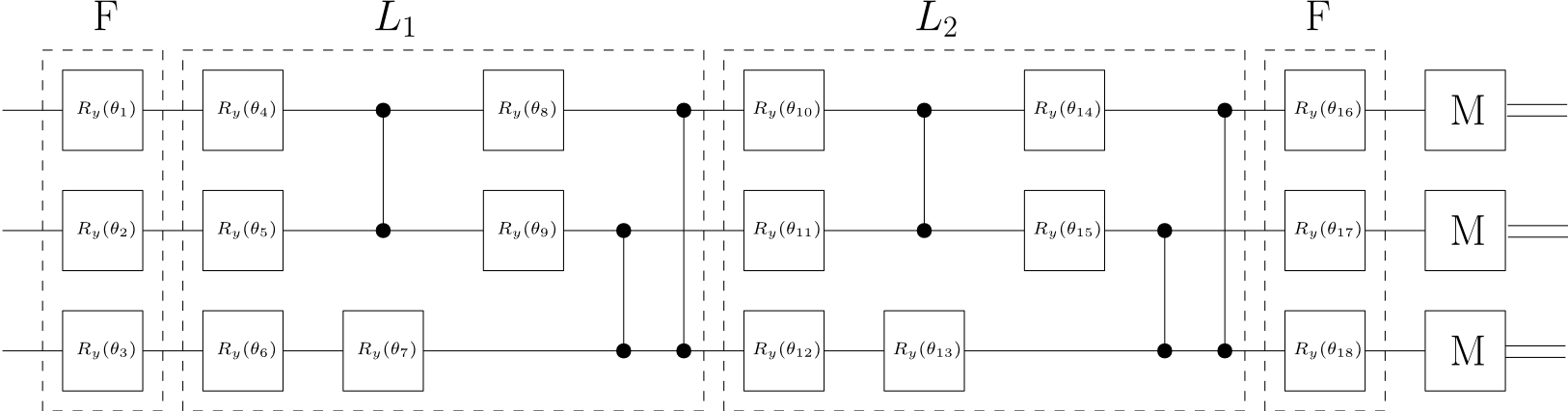} \\
		(B) \\
		\includegraphics[width=0.80\textwidth]{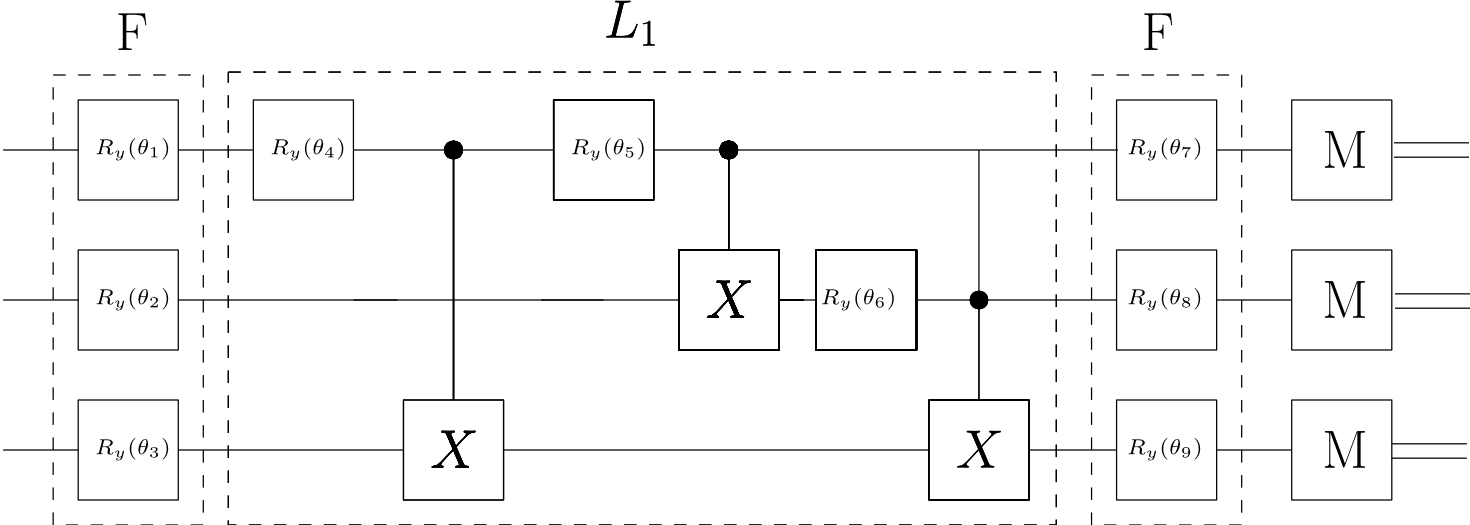} \\
		\end{tabular}	
\end{center}	
	\caption{Two ansatz circuits that are used in the credit data classification computational experiment. Circuit (A) is a typical ansatz circuit as known in the literature where rotation $Y$ gates and controlled-$Z$ gates are used. Its universal properties enable its wide range of applications for a different groups of data (also in the discussed issue). Circuit (B) is a simplified version adapted to the data discussed in this article. Its advantage is the smaller number of rotation gates $R_y$ and we use controlled-$NOT$ gate instead of controlled-$Z$ gates}
	\label{lbl:fig:ansatz:classification:circuits:JW:MS:ICCS:2023}
\end{figure}

Fig.~\ref{lbl:fig:ansatz:classification:circuits:JW:MS:ICCS:2023} illustrates two examples of ansatz circuits which were utilized to perform the bivalent classification in this article. Circuit (A) is a universal type of ansatz circuit \cite{Wecker2015} that is currently being proposed for the use in the VQE approach. The structure shown in the circuit (A) consists of two parts. In the first part, the first three rotation gates are responsible for encoding the state features, but as we have already mentioned in Sec.~\ref{lbl:ssec:entanglement:in:probes:iccs:2023:jw:ms}, the states describing the data are entangled, thus, the second part of the circuit is responsible for introducing entanglement. This is realized by means of controlled change of sign gates and rotation gates, which allow changing the level of entanglement. The number of repetitions of the second part is also important, because greater accuracy can be obtained by increasing the number of repetitions, i.e. by increasing the depth of the circuit. However, two layers, as it will be shown later, are sufficient to obtain a high level of correct classification.	

The variant (B) of the circuit was created to directly build states consistent with the general structure of the data encoded in the undertaken classification task. The circuit is characterized by less depth and only one layer gives high accuracy of classification results. This allowed us to built less complicated circuit with smaller number of rotation gates needed to implement the VQE circuit.

\subsection{VQE for Data Classification} \label{lbl:ssec:alg:for:data:class:JW:MW:ICCS:2023}

The algorithm of classification is presented in a form of pseudocode as Alg.~\ref{lbl:fig:classification:with:VQE:EC:2024:JW:MW}. We utilize nine-qubit register and its initial state may be written in three parts $\mket{\Psi}=\mket{\psi_{anc}}\mket{\psi_{VQE}}\mket{\psi_{D_i}}$, where the subregister $\mket{\psi_{VQE}}$ marks the qubits dedicated to the VQE, $\mket{\psi_{D_i}}$ represents $i$-th sample from data set $D$, and $\mket{\psi_{anc}}$ are auxiliary qubits necessary to realization of the SWAP-Test.

Generally, it is possible to apply various functions to calculate the error value. It can be realized by the function $f$, given in Eq.~(\ref{lbl:eq:f:error:function:JW:MS:ICCS:2023}), but in Alg.~\ref{lbl:fig:classification:with:VQE:EC:2024:JW:MW} it is performed directly by the SWAP-Test. The algorithm's input is a data set $\{ D \}$. Each observation is evaluated with the use of previously calculated parameters $\theta_{T^g_c}$. These parameters are the output of the optimization process as in Eq.~\ref{lbl:eq:VQE:optimalization:process} for each class $c$ and group $g$ (the data analysis showed that there are groups of points closely located to one another in each class what may be observed in Fig.~\ref{lbl:fig:pca:decomposition:two:comp:training:data}). 

The function $f$ and the SWAP-Test calculate a distance between a vector of analyzed observation $i$ and a class $c$. More precisely, both functions compute the distance between an observation $i$ and quantum state which is pointed out by parameters $\theta_{T^g_c}$. An acceptance threshold -- which is the boundary distance from the class center to its most distant elements -- should not be to high because it may cause the class overlapping. For the analyzed data set, the initial acceptance threshold is given by $\Omega^g_c=0.5$ and its lowest value is $\Lambda^g_c=0.0$. The optimization process should lead to the minimization of $\Omega^g_c$ to reduce the number of situations when one observation might be categorized to more than one class.      

It should be added that an observation recognized as an element of a class is placed in the set $C$ and non-categorized observations in the set $U$. Because values of the parameters $\theta_{T^g_c}$ are changed in the optimization process, the additional filtration is required for elements of the set $U$.

The classification algorithm is based on one instance of the VQE method but by changing parameters $\theta_{T^g_c}$ it may be reconfigured for each class $c$ and group $g$. Alg.~\ref{lbl:fig:classification:with:VQE:EC:2024:JW:MW} is characterized by a polynomial complexity:
\begin{equation}
T\left( \mket{\Psi}, \theta_{T^g_c}, D \right) = \sum_{i=0}^{N-1} \sum_{k=0}^{c_n - 1} \sum_{l=0}^{c^g_n - 1} \left( \underbrace{T_{ini}+T_{VQE} + T_{SWAP} + T_{oth}}_{T_{qtot}} \right) = O( N \cdot c_n \cdot  c^g_n \cdot T_{qtot}) ,
 \end{equation}
as a consequence of three $for$ loops ($T_{ini}$ represents a constant time necessary for the initialization of quantum part of the algorithm/circuit). Theoretically, the maximal complexity attains maximally $O(N^3)$ if the number of classes and groups is also equal to the total number of probes $N$. In the case of our data, we assume that we have $N$ probes and two main classes -- zero ($0$) and one ($1$) -- then the complexity can be expressed as $T(N \cdot 0^g_n 1^g_n)$ where the total number of probes is $N$ and the number of probes in groups for two classes have the influence on the complexity of classification computation task. The complexity of $T_{qtot}$ (the part realized by the quantum circuits) dependents only on the depth of $VQE$ circuit because the SWAP-Test has a constant depth of its circuit, equal to three layers. The depth of the VQE circuit, in general, should be regarded as a poly-logarithmic function $O(\log^c \frac{1}{\varepsilon})$ and $c \approx 4$. In this case, $\varepsilon$ is an accuracy of the obtained state after learning process of the VQE \cite{Bravo:Prieto:2020:Scaling:Of}. Due to the exponential complexity, the number of necessary qubits is always a logarithmic relation, for single probe with dimensionality $d$, we need only logarithmic number of qubits $q_n$:
\begin{equation}
	q_n = \log_2 d .
\end{equation}
In the analyzed problem, the eight input features require the use of three qubits. 

Additionally, an adequate number of the VQE instances may be called in parallel and observations may be distributed to independent circuits to realize a parallel processing what allows obtaining an additional linear speed-up. 	

\begin{algorithm}[!ht]
\SetAlgoNoLine
\caption{MVQEClassification( $\mket{\Psi}$, $\theta_{T^g_c}$, $D$ ) $\longrightarrow$ (C, U) \\
The algorithm of data classification for a set $D$ ($N$ symbolizes the number of observations). The whole quantum register is marked as $\mket{\Psi}$ and $\theta_{T^g_c}$ are parameters of the ansatz circuit for class $c$ and group $g$. For each class, the acceptance range $(\Lambda^g_c, \Omega^g_c)$ is given. Symbol $c_n$ denotes the number of classes and $c^g_n$ the number of groups in a class $c$ . Index $i_c$ means that i-th probe belongs to class $c$}
\label{lbl:fig:classification:with:VQE:EC:2024:JW:MW}

\KwIn{ $\mket{\Psi}$, $\{ \theta_{T^g_c} \}$, $D$ }
\KwOut{ $C$, $U$ }

$C=\emptyset$; $U=\emptyset$ \;
\For{$i \gets 0, \ldots, N-1$}{
	\For{ $c \gets  0,  \ldots, c_n - 1$}{
		\For{ $l \gets 0, \ldots, {c^g_n} - 1 $} {
				initiate subregister $\mket{\psi_{anc}}$ with $\mket{0}$ \;
				initiate subregister $\mket{\psi_{VQE}}$ with $\mket{0}$ \;
				initiate subregister $\mket{\psi_{D_{i}}}$ with $D_i$ \;
				perform VQE( $\mket{\psi_{VQE}}$, $\theta_{T^{g=l}_c}$) $\rightarrow$ $\mket{\phi^{g=l}_c}$ \;
				$P_{D_i}$ = perform  SWAP-Test$\left( \mket{\phi^{g=l}_c}, \mket{\psi_{D_{i}}} \right)$ \;
				\If{ $\Lambda^g_c <  P_{D_i} < \Omega^g_c$ }{
				$C_{i_c} = C_{i_c} \cup D_i $ \;  
				probe $\mket{\psi_{D_{i}}}$ is classified for class $c$ (group $g$) \;
				}\Else{
				$U = U \cup ( \mket{\psi_{D_{i}}}, c) )$ \; 
				probe $\mket{\psi_{D_{i}}}$ in not recognized as probe for class $c$ \;
				} 
			} 
		} 
} 
\textbf{return} C, U	
\end{algorithm}

\subsection{Accuracy of Presented Approach} \label{lbl:subsec:accuracy:JW:MS:ICCS:2023}

Validation of the presented solution was performed using several VQE variants. Table~\ref{lbl:tbl:vqe:accuracy:JW:MS:2023} represents the obtained results for the analyzed data set. The overall effectiveness, i.e. the number of correctly classified samples, was checked. In addition, in the case of the proposed solution, it was also verified how a given classifier deals with samples from the other class, which allows to indicate the cases of a false classification. It should also be noted that the VQE circuit generates a state that indicates a given class with a very low error rate. Calculated values of the cost function $f$, Eq.~(\ref{lbl:eq:f:error:function:JW:MS:ICCS:2023}), are less than $\approx 0.25$ and usually take a value in the range $(0.05, 0.1)$.    	

\begin{table}[!ht]
	\caption{The VQE classification quality for several variants of quantum circuits with different versions of the ansatz circuit. The $q$ parameter is the number of used qubits to encode features of observations, $f$ stands for the number of values for encoding features, $al$ is the number of the ansatz circuit values. The expression std-ansatz-tuned-epsilon means that the parameter $\Omega^g_c$ was manually tuned, e.g. by selecting acceptance thresholds and the learning process for the VQE was repeated several times. The name  std-tuned-ansatz-tuned-epison means that (B) variant ansatz was used and again epsilon parameters has been tuned for better sensitivity of the classifier}
	\begin{center}
		\begin{tabular}{|l||l|l|c|}
			\hline
			No. & Classifier short name						& Basic parameters 			& \parbox{2cm}{Accuracy} \\ \hline \hline
			1	& $VQE^8$									& q=8, f=1, al=1			& 	71.25\%  			 \\	\hline
			2	& $VQE^8$									& q=8, f=1, al=2			& 	81.25\%  			 \\	\hline
			3	& $VQE^8$									& q=8, f=1, al=3			& 	88.75\%  			 \\	\hline			
			4	& $VQE^8$									& q=8, f=2, al=2			& 	91.25\%  			 \\	\hline			
			5   & $VQE^3$, std-ansatz						& q=3,  al=1				& 	85.00\% (22\%)  	 \\	\hline			
			6   & $VQE^3$, std-ansatz						& q=3,  al=2				& 	95.00\% (27\%)   	 \\	\hline			
			7   & $VQE^3$, std-ansatz-tuned-epsilon			& q=3, al=2				& 	100.00\% (0\%)  	 \\	\hline			
			8   & $VQE^3$, std-tuned-ansatz					& q=3,  al=1				& 	87.50\% ( 32\%)   	 \\	\hline			
			9   & $VQE^3$, std-tuned-ansatz					& q=3,  al=2				& 	97.50\% ( 7\%)   	 \\	\hline			
			10  & $VQE^3$, std-tuned-ansatz-tuned-epsilon	& q=3,  al=1				& 	98.50\% ( 1\%)   	 \\	\hline			
			11  & $VQE^3$, std-tuned-ansatz-tuned-epsilon	& q=3,  al=2				& 	100.0\% ( 0\%)   	 \\
			\hline			
		\end{tabular}
	\end{center}		
	\label{lbl:tbl:vqe:accuracy:JW:MS:2023}
\end{table}

The best result was achieved, for the approach described in this paper, by two ansatz circuits depicted in Fig.~\ref{lbl:fig:ansatz:classification:circuits:JW:MS:ICCS:2023}.  Especially, the case (B) where, in addition to training the circuit several times, the parameters of the classification threshold were also properly selected, what provided a high, 100\%, classification efficiency (No. 7 and 11, Table~\ref{lbl:tbl:vqe:accuracy:JW:MS:2023}) by only two-layer circuit. The version (B) can also get 100\% accuracy with only one layer after tuning $\Omega^g_c$ values. All samples were correctly classified and there were no false indications in class 1. It should be also noticed that the use of three qubits, to describe the samples, also speeds up the optimization process. The learning time for the SPSA optimizer \cite{Spall1997}, \cite{Wang2018}, \cite{Agafonov2021} with 500 iterations was about 30 (the worst case 35) seconds, where Intel Xeon W-2245 3.9 Ghz processor was used, OS Windows 11, and Python 3.10.7. For comparison, in case of the VQE method based on eight qubits, the learning time was over 600 (the worst time was over 700 for the case No. 4, Table~\ref{lbl:tbl:vqe:accuracy:JW:MS:2023}) seconds for the COBYLA optimizer \cite{Powell1994}, \cite{Arrasmith2020}. Nevertheless, high classification efficiency was also achieved here, and increasing number of the ansatz circuit layers, as can be seen, translates into increased accuracy.


\section{Conclusions} \label{lbl:sec:conclusions:iccs:2023:jw:ms}

In this work, we have shown that the VQE approach together with the SWAP-Test may be utilized to perform the classification task. Firstly, we have normalized the analyzed data set. This process ensured avoiding excessive influence of some variables on the constructed model and transformed the classical observations to quantum states. This strategy of data preparation is universal and may be applied for variables of any type.

Using the classical method of machine learning, we have shown in Sec.~\ref{lbl:sec:accuracy:of:classical:techniques:iccs:2023:jw:ms} that constructed decision problem may be solved by the classical computer with a high accuracy. However, the classical models which were capable to fit to the analyzed data were quite complex (e.g. SVM with a polynomial kernel function or Random Forest Classifier).

In the presented approach, which combines the VQE method and the SWAP-Test, it may be seen that the solution is simple and requires a logarithmic complexity of constructed circuits because of the natural feature of quantum states -- their exponential capacity.

During the experiments the presence of entanglement was detected in the prepared samples. The VQE method also allowed us to model the classes with preservation of the entanglement phenomenon. The source code of all conducted numerical experiments, including plots generation, may be found in the code repository \cite{VQEClassRepo}.

%
%
\section*{Acknowledgments} \label{lbl:sec:acknowledgments:2024:jw:ms}

We would like to thank for useful discussions with the~\textit{Q-INFO} group at the Institute of Control and Computation Engineering (ISSI) of the University of Zielona G\'ora, Poland. We would like also to thank to anonymous referees for useful comments on the preliminary version of the paper. The numerical results were done using the hardware and software available at the ''GPU $\mu$-Lab'' located at the Institute of Control and Computation Engineering of the University of Zielona G\'ora, Poland.

\bibliographystyle{plain}
\bibliography{sci-art-VQE-credit-risk}

\end{document}